# Flowgraph Models and Analysis for Markov Jump Processes

BY MUHAMMAD FIKRI BUDIANA

*School of Mathematics, Faculty of Engineering Physical Science, The University of Manchester, Oxford Road, Manchester, M13 9PL, U.K.*

fikribudiana@gmail.com

AND MURWAN H. M. A. SIDDIG

*School of Mathematics, Faculty of Engineering Physical Science, The University of Manchester, Oxford Road, Manchester, M13 9PL, U.K.*

murwan_siddig@hotmail.com

*Abstract*—Flowgraph models provide an alternative approach in modeling a multi-state stochastic process. One of the most widely used stochastic processes that have many real-world applications especially in actuarial models is the Markov jump process or continuous- time Markov chain. However, finding waiting time distributions between any two states in a Markov jump process can be very difficult. Flowgraph analysis for Markov jump process comprises of modeling the possible states of the process, the interstates waiting time distribution, and working on the moment generating function domain to obtain the total waiting time distribution in form of density or survival function. This paper gives the theory and computational method of flowgraph analysis, uses it in Markov process problems, and compares the traditional Markov process construction method with the flowgraph method to demonstrate the convenience and practicality of flowgraph analysis.

## I. INTRODUCTION

FLOWGRAPH models are one specific form a multistate model, a model that is used to illustrate time-to-event data resulting from a stochastic process. The stochastic processes that are modeled into a multi-state model have several possible states or outcomes and they are usually used to explain the movement of an individual that progress through those states but can only attend one state at a time. For example, in the calculation of insurance premium which pays out if the insured is ill or dead, the states of the process are the state of the body of the insured, whether he is healthy, ill, or dead. The main interests are usually in the probability or time of transition between states and the expected waiting time in each state until the next transition.

A flowgraph is made by nodes serving as the possible outcomes or states, in which the nodes are connected by branches. Each branch is a directed line segment labeled with the corresponding transition probability and waiting time distribution, which is represented by the moment generating function (MGF). The main purpose of flowgraph analysis is to calculate or predict the waiting time distribution between any two nodes of interest in a flowgraph. The output of a flowgraph analysis is the MGF of the waiting time of interest, which can be converted into the density, cumulative distribution function (CDF), or survival and hazard function of the waiting time. Obtaining the waiting time distribution between two states in a multi-state model is very important for further calculation and data analysis of a multi-state model. For example, Loeffen (2014) shows that one quantity that plays an important role in the computation of expected present value of an insurance premium is the risk of dying, i.e. the waiting time distribution from healthy state to the death of the insured.

The most comprehensive explanation of flowgraph model and its usefulness for the analysis of time-to-event data are presented in the book of Huzurbazar (2005), which will be the main reference for this report. This enhancement extends the use and flexibility of flowgraph models.

One of the mainly used multi-state models is the continuous- time Markov chain, also known as the Markov jump process (MJP). MJP models the progression of a stochastic process through a finite number of possible states in a continuous time space. It satisfies Markov property and focuses on the transition rates between the states, the transition probabilities, as well as the waiting time in each state. The sample path of an MJP can be constructed to be used as a tool for analyzing a multi-state model.

To establish notation, we let $T \in [0, \infty]$ be the time to event. We assume $f(t)$ be the density of $T$ and $F(t)$ be the cumulative distribution function (CDF) of $T$. The survival function of $T$ is defined to be is

$$S(t) \coloneqq P(T > t) = 1 - F(t), \quad (1)$$

and the hazard function is

$$h(t) := \lim_{h \downarrow 0} \frac{1}{h} P(T \leq t+h \mid T \geq t)$$

$$= \lim_{h \downarrow 0} \frac{1}{h} \frac{(P \leq T \leq +h)}{P(T \geq t)}$$

$$= \frac{1}{S(t)} \lim_{h \downarrow 0} \frac{F(t+h) - F(t)}{h} \quad (2)$$

$$= \frac{1}{S(t)} \lim_{h \downarrow 0} \frac{\int_t^{t+h} f(s)ds}{h} = \frac{f(t)}{S(t)}.$$

## II. FLOWGRAPH MODELS

Figure 1 is an example of a flowgraph model, a general reversible illness–death model. A person started from a healthy state denoted by state 0. The healthy person in state 0 can either transition to state 1 or advance directly to state 2. State 1 is a diseased state and state 2 represents death. Once diseased, the person can get a treatment and become healthy again, moving back to state 0, thus giving the name 'reversible'.

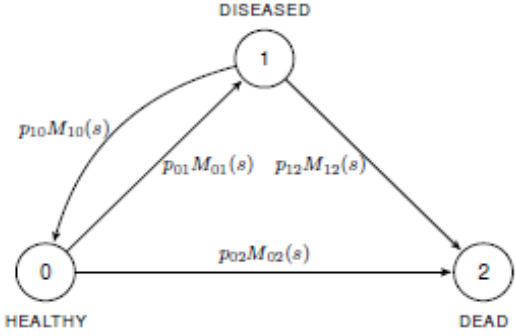

Figure 1. Flowgraph model for a general reversible illness-death model

The branches are labeled by the transition probabilities and moment generating functions associated with the waiting time distributions for each transition to occur.

*Definition 1:* The moment generating function (MGF) of a random variable $T$ is defined by

$$M_T(s) = E(e^{sT}) = \int_{-\infty}^{+\infty} e^{st} f_T(t) \, dt, \quad (3)$$

provided that $M_T$ exists for $s \in (-a, a)$ and $a > 0$ The output a flowgraph is the MGF of the waiting time distribution of interest. In this example, the waiting time of interest can be the waiting time of a healthy person until his death due to a disease, $0 \to 1 \to 2$, or it may be the waiting time from healthy to dead state without concerning which path is taken, $0 \to 2$.

Flowgraph models are outcome graphs, which mean that their states or nodes represent possible outcomes. The states are connected by edges or lines, where the edges have a direction associated with them. Those directed edges are called *branches*. Every branch has its own transition probability and waiting time distribution associated with the transition it represented. The labels on the branches are called *transmittance,* which contain probabilities and MGFs.

*Definition 2:* The transmittance consists of the transition probability times the MGF of the distribution of the waiting time.

In Figure 1, the transition from healthy to diseased state $0 \to 1$ has transmittance $p_{01}M_{01}(s)$. The transition probability is $p_{01}$, the probability of transition to get to a disease for a healthy person. $M_{01}(s)$, the MGF, represents the waiting time distribution in state 0 until transition to state 1. Branch transmittances of a flowgraph model are useful for solving the distribution of the waiting time of interest.

*Definition 3*: The overall transmittance refers to the transmittance of the whole flowgraph from the initial until the final state.

### A. Series flowgraph structure

The most elementary component of a flowgraph is the simple series system. In a series flowgraph, the only allowable transitions are progressions from one state to the next state. That transition occurs with certainty, such that is, the transition probability is equal to 1.

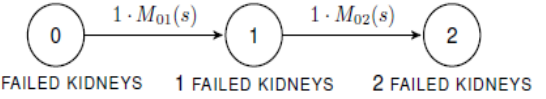

Figure 2. Series flowgraph for kidney failure

*Example 1 (Simple Series Model):* Figure 2 shows a model of kidney failure with three states. State 0 represents the patient in early diagnosis of kidney disease, where two of his kidneys are still functioning. State 1 represents one failed kidney, and state 2 represents both of his kidney have stop functioning. Suppose that $Y_0$ represents the random waiting time in state 0 until one kidney fails and state 1 is reached, and $Y_1$, independent of $Y_0$, represents the random waiting time in state 1 until the second kidney fails and state 2 is reached. The waiting time of interest is the survival time of the patient's kidneys, the total waiting time from $0 \to 2$. Let $T = Y_0 + Y_1$ be that total waiting time. In this flowgraph, the transition probabilities are $p_{01} = p_{12} = 1$ since each state will certainly progress to the next state. The MGFs corresponding to the waiting time distribution of each transition are $M_{01}(s) = E(e^{sY_0})$ and $M_{12}(s) = E(e^{sY_1})$. The transmittances, $p_{01}M_{01}(s)$ and $p_{12}M_{12}(s)$, are written on the branch of the flowgraph.

### B. Parallel flowgraph structures

Another basic element of a flowgraph is the parallel system, where the branches are in parallel with each other. In a parallel flowgraph, one beginning state can progress to one of the states from a set of possible outcomes. This situation is called competing risks.

*Example 2 (Medical Application: Progression of Cancer):* Figure 3 presents a flowgraph model for the progression of cancer. State 0 is the initial state of cancer, state 1 is the advanced stage, and state 2 represents death. In this example, the parallel states are states 1 and 2.

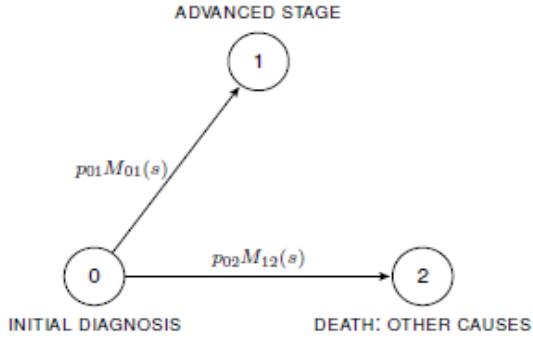

Figure 3. Parallel flowgraph for cancer progression

The transition probabilities from state 0 to the next stages are $p_{01}$ and $p_{02}$, where $p_{01} + p_{02} = 1$, with the MGFs of the corresponding waiting time distributions are $M_{01}(s)$ and $M_{02}(s)$. Let $Y_1$ be the waiting time to state 1 and $Y_2$ be the waiting time to state 2. From the flowgraph structure, we can deduce that the overall waiting time is $\min(Y_1, Y_2)$ If we assume that $Y_1$ and $Y_2$ follow some distribution, then the transition probability to state 1 is $p_{01} = P(Y_1 < Y_2)$.

To model the competing risk in a flowgraph, we use a conditional approach. The probability is given by

$$P[\min(Y_1, Y_2) \leq t] = P(Y_1 \leq t|0 \to 1)p_{01} \\ + P(Y_2 \leq t|0 \to 2)p_{02} \\ = P(Y_1 \leq t|Y_1 \leq Y_2)P(Y_1 < Y_2) \\ + P(Y_2 \leq t|Y_2 \leq Y_1)P(Y_2 < Y_1) \quad (4)$$

### C. Combinations of series and parallel flowgraph

There will be more than one possible path from the initial to the final state.

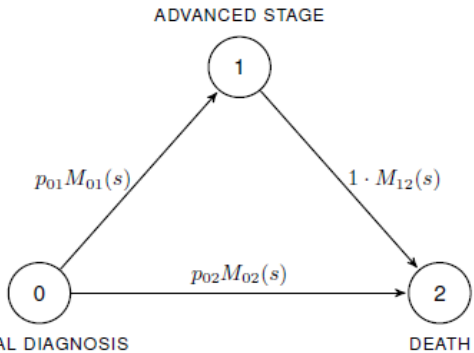

Figure 4. Flowgraph model for cancer progression

*Definition 4*: A path from state $i$ to state $j$ is any sequence of nodes that starts from $i$ and ends in $j$ that does not go through any intermediary states more than once.

*Example 3 (Medical Application: Progression of Cancer):*
Figure 4 presents a more complex flowgraph model for progression of lung cancer. State 0 represents the initial diagnosis of cancer, state 1 represents an advanced stage of cancer, and state 2 represent the patient's death. States 0, 1, 2 are in series and states 1 and 2 are in parallel. The quantities of interest are time from the initial diagnosis until death, whether it is due to any causes, advanced cancer, or other causes except for cancer.

### D. Loop flowgraph model

Feedback loop is the third basic component of a flowgraph after the series and parallel elements.
*Definition 5:* A loop is a path whose endpoint is the same as the initial state.

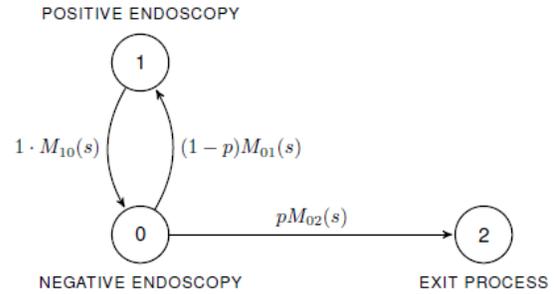

Figure 5. Flowgraph model for ulcer recurrence

*Example 4 (Recurrence of an Ulcer):* Figure 5 shows a loop flowgraph model of the recurrence of an ulcer. State 0 is the initial state for a patient diagnosed with ulcer that has received a therapy to heal it. To detect an ulcer recurrence, patient goes through an endoscopy procedure. If the result is positive, it indicates that the ulcer has recurred, and the patient progress to state 1. The patient is then treated and goes back to state 1. If the ulcer does not recur, the patient exits the process and ended up in state 2. The feedback loops are $0 \to 1 \to 0$ and $1 \to 0 \to 1$.

### E. Solving flowgraph models

Solving a flowgraph model is the act to reduce all of the branch transmittances to only a single branch with one overall transmittance.
*Definition 6:* An equivalent transmittance, denoted by $T_{(s)}$ is a transmittance which is attained after two or more branch transmittances are reduced into one transmittance.

### F. Solving series flowgraphs

Example 1, the kidney disease progression, is a simple Series system. The corresponding Figure 2 can be solved by computing the overall transmittance, the transmittance of path $0 \to 1 \to 2$.
*Definition 7*: The term path transmittance refers to the multiplication of every branch transmittances of the corresponding path.
The total waiting time $T$ has a distribution that equals to the distribution of the sum of those two independent waiting times that is $T = Y_0 + Y_1$. The MGF of $T$ can be expressed as

$$\begin{aligned} M_T(s) &= M_{Y_0+Y_1}(s) \\ &= E(e^{s(Y_0+Y_1)}) \\ &= E(e^{sY_0})E(e^{sY_1}) \\ &= M_{Y_0}(s)M_{Y_1}(s) \\ &= M_{01}(s)M_{12}(s) \end{aligned}$$

We can then build an equivalent flowgraph by removing node 1 and passages connecting it to the other nodes, so there is only one branch from state 0 to state 2 which is labeled by the overall transmittance $M_{01}(s)M_{12}(s)$. Figure 6 below is the said equivalent flowgraph.

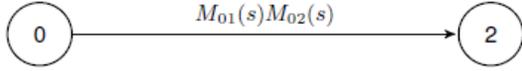

Figure 6. Solved flowgraph for a series structure

*Example 5 (Kidney Disease Progression)*: This example is based on the Example 1 but with the additions of distributional assumptions for the model. Assume that each kidney fails independently and following the exponential distribution with mean $1/\lambda_0$.
Suppose that $W_1$ and $W_2$ be the waiting time for the first and second kidneys to fail respectively. Then $W_1$ and $W_2$ are independent and identically distributed (i.i.d.) with $\text{Exp}(\lambda_0)$ distribution. The density and CDF of $W_1$ are

$$f(w) = \lambda_0 e^{-\lambda_0 w}, w > 0, \lambda_0 > 0 \quad (6)$$
$$F(w) = 1 - e^{-\lambda_0 w}, w > 0, \lambda_0 > 0 \quad (7)$$

We also assume that once one kidney fails, the remaining kidney's failure time is now follows an exponential distribution with a new parameter $\lambda_1$ such that $\lambda_1 > \lambda_0$.
The waiting times for the first and second kidneys to fail. Let $W_{(1)}$ be that minimum waiting time. Then $W_{(1)} = \min(W_1, W_2)$ is the minimum of two independent exponential distributions.
The distribution of $W_{(1)}$ is computed as follows:

$$\begin{aligned} P(W_{(1)} \le t) &= P(\min(W_1, W_2) \le t) \\ &= 1 - P(\min(W_1, W_2) > t) \\ &= 1 - P(W_1 > t, W_2 > t) \\ &= 1 - P(W_1 > t)P(W_2 > t) \quad (8)\\ &= 1 - [1 - F(t)]^2 \\ &= 1 - [1 - (1 - e^{-\lambda_0 t})]^2 \\ &= 1 - e^{-2\lambda_0 t}. \end{aligned}$$

Since the CDF of $W_{(1)}$ is $1 - e^{-2\lambda_0 t}$, we can conclude that it follows an exponential distribution with parameter $2\lambda_0$. Now we let $Y_0 = W_{(1)}$ be the waiting time from $0 \to 1$ and $Y_1$, independent of $Y_0$, be the waiting time from $1 \to 2$. The total waiting time for passage from state 0 to state 2 is distributed as the sum of the two independent exponential distribution, $T = Y_0 + Y_1$. The MGF of an exponential distribution with parameter $\lambda$ is given by $M(s) = \lambda/(\lambda - s)$ for $s < \lambda$. From (5) we have that

$$\begin{aligned} M_T(s) &= M_{Y_0+Y_1}(s) \\ &= M_{01}(s)M_{12}(s) \\ &= \frac{2\lambda_0}{2\lambda_0-s} \frac{\lambda_1}{\lambda_1-s} \quad \text{for } s < \min(2\lambda_0, \lambda_1). \end{aligned} \quad (9)$$

Figure 7 below is the equivalent flowgraph of Figure 2 labeled with the overall transmittance $\frac{2\lambda_0}{2\lambda_0-s} \frac{\lambda_1}{\lambda_1-s}$,

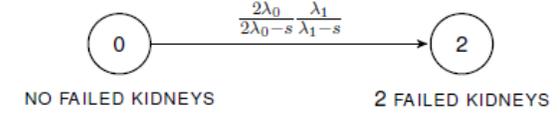

NO FAILED KIDNEYS    2 FAILED KIDNEYS

Figure 7. Solved flowgraph for kidney disease progression

*Corollary 1 (General results for convolution)*: Let $Y_0, Y_1, \ldots, Y_n$ be independent random waiting times such that $Y_i$ is the waiting time from $(i-1) \to i$ for all $i$ between 1 and $n$. Thus the MGF of the total waiting time $T = \sum_{i=0}^{n-1} Y_i$ is

$$M_T(s) = \prod_{i=0}^{n-1} M_{Y_i}(s), \quad (10)$$

with $M_{Y_i}(s)$ be the MGF of $Y_i$.

### G. Solving parallel flowgraphs

In a parallel flowgraph, the overall waiting time is the minimum of the waiting times from the input to the multiple possible outputs. Solving Example 2, The passage $0 \to 1$ has transition probability $p_{01}$, MGF $M_{01}(s)$, and hence the branch transmittance is $p_{01}M_{01}(s)$. Similarly, the passage from state 0 to 2 has probability of transition $p_{02} = 1 - p_{01}$, MGF $M_{02}(s)$, and branch transmittance $p_{02}M_{02}(s)$. Following (4), the MGF of the minimum waiting time to either state is then computed as

$$M_{(s)} = M_{\min(Y_1,Y_2)}(s) = p_{01}M_{01}(s) + p_{02}M_{02} \quad (11)$$

In general, for any $(n + 1)$-state parallel flowgraph in which the transition is possible from state 0 to states 1, 2... or n, the MGF of the overall waiting time distribution is

$$M(S) = \sum_{j=1}^{n} p_{0j}M_{0j}(s) \quad \text{where } \sum_{j=1}^{n} p_{0j} = 1. \quad (12)$$

### H. Solving combinations of series and parallel flowgraph

Example 3 presented a combination of series and parallel flowgraph model for cancer progression. We can see from Figure 4 that the upper path $0 \to 1 \to 2$ makes up a series structure. So to solve the whole flowgraph, we need to find the transmittance for that upper path first. Therefore the transmittance for the path $0 \rightarrow 1 \rightarrow 2$ is $p_{01}M_{01}(s)M_{12}(s)$ where $p_{12} = 1$. The flowgraph of Figure 4 can be replaced by the reduced flowgraph in Figure 8, a flowgraph with two parallel paths going directly to node 2 from node 1.

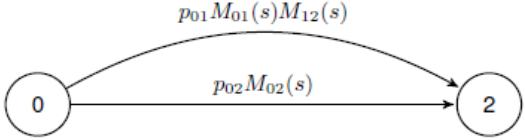

Figure 8. Reduced flowgraph model for a closed parallel system

The reduced flowgraph above can be dealt with by considering it as a parallel system with two branches. Using the equation (11), the overall MGF is computed as

$$M(s) = p_{01}M_{01}(s)M_{12}(s) + p_{02}M_{02}(s) \quad (13)$$

Therefore the flowgraph can be reduced once more to the solved equivalent flowgraph in Figure 9 with only one branch labeled with the equivalent transmittance $M(s) = p_{01}M_{01}(s)M_{12}(s) + p_{02}M_{02}(s)$.

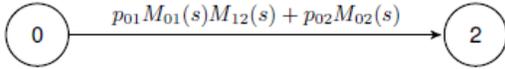

Figure 9. Solved flowgraph model for a closed parallel system

*Example 6 (Medical Application: Progression of Cancer):* We consider Example 3. Let $W_1$ and $W_2$ be the waiting times from state 0 to state 1 and state 2 respectively. We suppose that $W_1$ is exponentially distributed with parameter $\lambda_1$ and independent of $W_2$, which follows exponential distribution with parameter $\lambda_2$. The probability that the transition to state 1 occurs before the transition to state 2, $p_{01}$, is computed as follows:

$$\begin{aligned} p_{01} = P(W_1 < W_2) &= \int_0^\infty \int_{w_1}^\infty f_{w_1,w_2}(w_1, w_2) dw_2 dw_1 \\ &= \int_0^\infty \int_{w_1}^\infty \lambda_1 \lambda_2 e^{-\lambda_1 w_1} e^{-\lambda_2 w_2} dw_2 dw_1 \\ &= \int_0^\infty \lambda_1 e^{-\lambda_1 w_1} (-e^{-\lambda_2 w_2})|_{w_1}^\infty dw_1 \\ &= \int_0^\infty \lambda_1 e^{-(\lambda_1+\lambda_2)w_1} dw_1 \\ &= -\frac{\lambda_1}{\lambda_1 + \lambda_2} e^{-(\lambda_1+\lambda_2)w_1}\big|_0^\infty \\ &= \frac{\lambda_1}{\lambda_1+\lambda_2}. \end{aligned} \quad (14)$$

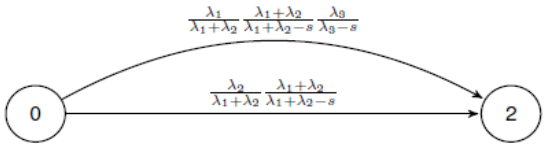

Figure 10. Reduced flowgraph model for cancer progression

The probability for direct passage to state 2 is then given by

$$p_{02} = P(W_2 < W_1) = 1 - p_{01} = \frac{\lambda_2}{\lambda_1 + \lambda_2}.$$

To compute the MGFs

$$\begin{aligned} P(W_1 \leq t \,|W_1 < W_2) &= \frac{P[W_1 \leq \min(t, W_2)]}{P(W_1 < W_2)} \\ &= \frac{\lambda_1 + \lambda_2}{\lambda_1} \int_0^t \int_{w_1}^\infty \lambda_1 \lambda_2 e^{-\lambda_1 w_1} e^{-\lambda_2 w_2} dw_2 d w_1 \\ &= \frac{\lambda_1 + \lambda_2}{\lambda_1} \int_0^t \lambda_1 e^{-\lambda_1 w_1}(-e^{-\lambda_2 w_2})|_{w_1}^\infty dw_1 \\ &= (\lambda_1 + \lambda_2) \int_0^t e^{-(\lambda_1+\lambda_2)w_1} dw_1 \\ &= -e^{-(\lambda_1+\lambda_2)w_1}|_0^t = 1 - e^{-(\lambda_1+\lambda_2)t} \quad \text{for } t > 0. \quad (15) \end{aligned}$$

We can see that it follows exponential distribution with parameter $\lambda_1 + \lambda_2$. The density function and MGF of the corresponding competitive waiting time distribution are then given by

$$\begin{aligned} f_{W_1|W_2<W_2}(t) &= (\lambda_1 + \lambda_2)e^{-(\lambda_1+\lambda_2)t} \\ M(s) &= \frac{\lambda_1+\lambda_2}{\lambda_1+\lambda_2-s} \quad \text{for } t > 0. \end{aligned}$$

Similarly, we can derive the competitive waiting time distribution of the direct passage to state 2 to occur first and we will have the same answer from (15).

Now we assume that waiting time from state 1 to state 2, $W_3$, follows $\text{Exp}(\lambda_3)$ distribution and is independent of $W_1$ and $W_2$. We have known how to solve this kind of flowgraph from the beginning of this subsection. First, solving the upper path $0 \rightarrow 1 \rightarrow 2$ will give the path transmittance:

$$\begin{aligned} &p_{01}M_{01}p_{12}M_{12}(s) \\ &= \left(\frac{\lambda_1}{\lambda_1+\lambda_2}\right)\left(\frac{\lambda_1+\lambda_2}{\lambda_1+\lambda_2-s}\right)\left(\frac{\lambda_3}{\lambda_3-s}\right) \end{aligned}$$

where $p_{12} = 1$. The transmittance of the direct passage from $0 \rightarrow 2$ is given by

$$p_{02}M_{02}(s) = \left(\frac{\lambda_2}{\lambda_1+\lambda_2}\right)\left(\frac{\lambda_1+\lambda_2}{\lambda_1+\lambda_2-s}\right).$$

The original flowgraph can be replaced by Figure 10 that has two parallel branches from state 0 directly to state 2. The waiting time distribution in state 0 is now a mixture of two distributions: the convolution of two independent exponential distribution $\text{Exp}(\lambda_1+\lambda_2)$ and $\text{Exp}(\lambda_3)$ with probability $\lambda_1/(\lambda_1+\lambda_2)$, and $\text{Exp}(\lambda_1+\lambda_2)$ with probability $\lambda_2/\lambda_1+\lambda_2$. Therefore, the overall transmittance for the solved equivalent flowgraph (shown in Figure 11) is given by

$$\begin{aligned} M(s) = & \left(\frac{\lambda_1}{\lambda_1+\lambda_2}\right)\left(\frac{\lambda_1+\lambda_2}{\lambda_1+\lambda_2-s}\right)\left(\frac{\lambda_3}{\lambda_3-s}\right) + \left(\frac{\lambda_2}{\lambda_1+\lambda_2}\right)\left(\frac{\lambda_1+\lambda_2}{\lambda_1+\lambda_2-s}\right), \\ & \text{for } s < \min(\lambda_1+\lambda_2, \lambda_3). \quad (16) \end{aligned}$$

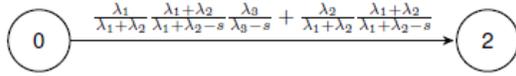

Figure 11. Solved flowgraph model for cancer progression

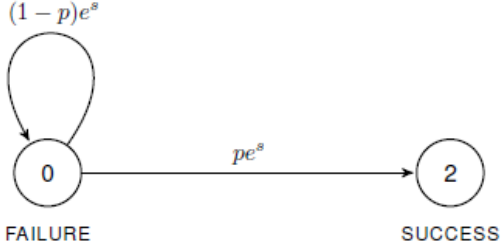

Figure 12. Flowgraph model for geometric distribution

### I. Solving flowgraphs with feedback loops

Feedback loop is the third main component of a flowgraph model after the series and parallel elements. It has a strong connection with a geometric distribution.

*Example 7 (Medical Example: Heartburn):* Figure 13 presents a flowgraph model for a cycle of heartburn. A patient starts from state 0 when he gets a heartburn condition. He may drink some drugs to get a temporary relief, transitioning him to the state $R$. He will then return to state 0 because the relief is only momentary, and the medicine will be taken again for him to get another temporary relief. After several cycles, the patient will progress to state 1 in which the heartburn cycle ended. Suppose that T is the total waiting time of the heartburn cycle. The states $0 \to R \to 0$ are in a series structure so it can be reduced to a feedback loop. For simplicity, we let

$$M_{00}(s) = M_{0R}(s)M_{R0}(s).$$

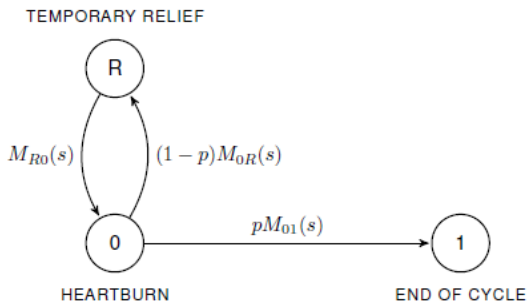

Figure 13. Flowgraph model for heartburn

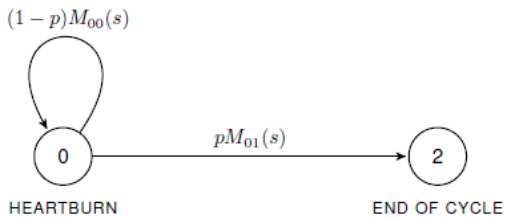

Figure 14. Partially reduced flowgraph model for heartburn

The reduced flowgraph is presented in the Figure 14 with the equivalent transmittance of the feedback loop equals to

$$(1-p)M_{00}(s) = (1-p)M_{0R}(s)M_{R0}(s).$$

Starting from state 0, the heartburn condition can directly end, making a transition into state 1. The MGF of the overall waiting time distribution for this case will be $pM_{01}(s)$. If the patient only experienced a temporary relief once before the cycle ended, the path will be $0 \to 0 \to 1$ and so the overall waiting time MGF will be $pM_{01}(s)(1-p)M_{00}(s)$. If the feedback loop is taken two times, the total waiting time MGF would be $pM_{01}(s)[(1-p)M_{00}(s)]^2$, and three times becomes $pM_{01}(s)[(1-p)M_{00}(s)]^3$. Doing the iteration infinite times will give the overall MGF:

$$\begin{aligned} M_T(s) &= pM_{01}(s) + pM_{01}(s)(1-p)M_{00}(s) \\ &\quad + pM_{01}(s)[(1-p)M_{00}(s)]^2 \\ &\quad + pM_{01}(s)[(1-p)M_{00}(s)]^3 + \cdots \\ &= pM_{01}(s)\{1 + (1-p)M_{00}(s) + \\ &\quad [(1-p)M_{00}(s)]^2 + \cdots\} \\ &= pM_{01}(s)\sum_{j=0}^{\infty}[(1-p)M_{00}(s)]^j \\ &= pM_{01}(s)\frac{1}{1-(1-p)M_{00}(s)} \text{ for } |(1-p)M_{00}(s)| < 1. \end{aligned}$$
(17)

We can also solve for the overall MGF by think of it as a parallel flowgraph with infinite parallel branches.

### J. Combining series, parallel, and loop flowgraphs

One example of a combination of series, parallel, and loop flowgraphs is the general reversible illness–death model, shown in Figure 15.

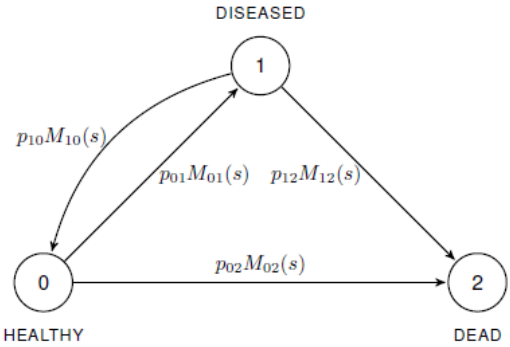

Figure 15. Flowgraph model for a general reversible illness–death model

We are interested in the waiting time from healthy to death regardless of which path is taken in between. We need to work on the lower path, $0 \to 2$, and upper path, $0 \to 1 \to 2$, separately to solve this problem. The lower path in this case is not just the direct passage from state 0 to state 2. It also involves the feedback loop $0 \to 1 \to 0$. Figure 16 represents the lower path and it can be reduced to Figure 17 by reducing the feedback loop. The equivalent transmittance for the

lower path can be computed using (17) to give

$$p_{02}\left[\frac{M_{02}(s)}{1-p_{01}p_{10}M_{01}(s)M_{10}(s)}\right] \quad (18)$$

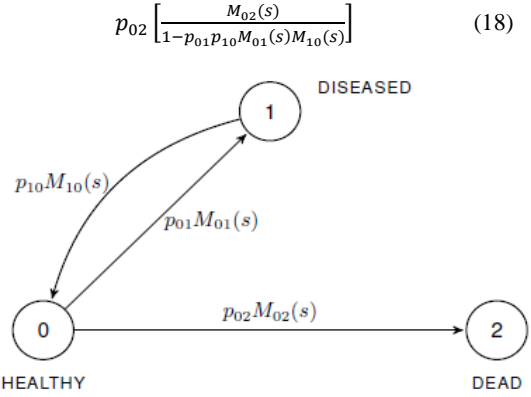

Figure 16. Subflowgraph for the lower path of Figure 15

Therefore the probability of taking the lower path is $p_{02}/(1-p_{01}p_{10})$, the value of the transmittance at $s = 0$. We can write the equivalent transmittance in the form of the probability of taking the path multiplied by the MGF of the lower path like this:

$$p_{02}\left[\frac{M_{02}(s)}{1-p_{01}p_{10}M_{01}(s)M_{10}(s)}\right] = \left(\frac{p_{02}}{1-p_{01}p_{10}}\right)\left[\frac{(1-p_{01}p_{10})M_{02}(s)}{1-p_{01}p_{10}M_{01}(s)M_{10}(s)}\right] \quad (19)$$

Figure 18 presents the subflowgraph for the upper path of reversible illness–death model. First we reduce the feedback loop into state 1 as shown in Figure 19. We can reduce it further by removing the feedback loop into just one branch from state 1 to state 2 and we will get a series structure like in the Figure 20. The equivalent transmittance for the upper path is then computed as

$$p_{01}\left[\frac{p_{12}M_{01}(s)M_{12}(s)}{1-p_{01}p_{10}M_{01}(s)M_{10}(s)}\right].$$

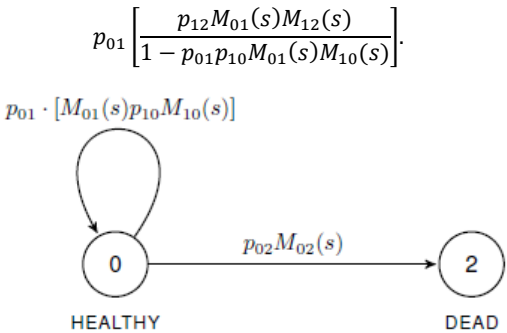

Figure 17. Reduced lower path of Figure 15

Similarly, with the lower path, we can write the transmittance as the probability of taking upper path multiplied by the MGF of upper path like this:

$$\left(\frac{p_{01}p_{12}}{1-p_{01}p_{10}}\right)\left[\frac{(1-p_{01}p_{10})M_{01}(s)M_{12}(s)}{1-p_{01}p_{10}M_{01}(s)M_{10}(s)}\right].$$

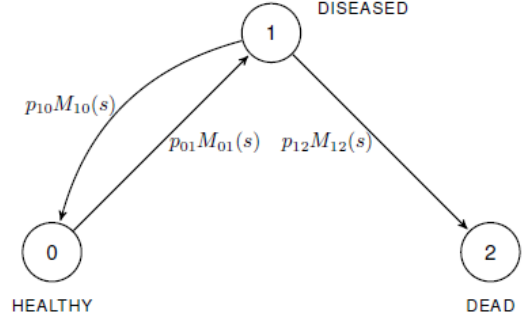

Figure 18. Subflowgraph model for the upper path of Figure 15

Both of the reduced lower and upper path can be joined to give a parallel flowgraph in Figure 21. We can now solve for the overall waiting time MGF for the reversible illness–death model just like solving a parallel flowgraph model with two branches:

$$M(s) = \frac{p_{01}p_{12}M_{01}(s)M_{12}(s)+p_{02}M_{02}(s)}{1-p_{01}p_{10}M_{01}(s)M_{10}(s)}. \quad (20)$$

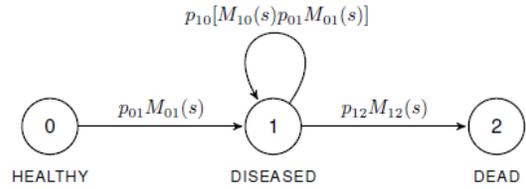

Figure 19. Reducing the feedback loop for the upper path

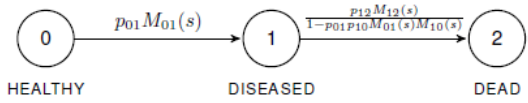

Figure 20. Reducing the feedback loop for the upper path

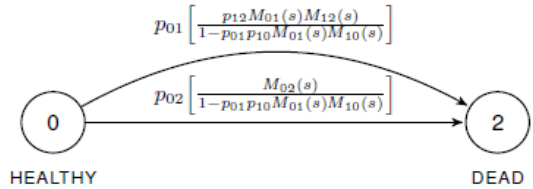

Figure 21. Reduced flowgraph model for Figure 15

K.  *Exact inversion of flowgraph MGF*

After getting the overall MGF of a flowgraph model, our interest is to convert it into density function. There are several methods to convert MGFs into density function. In this paper, we will only give examples of the exact inversion method in which it can only be used if the underlying waiting time distributions follow some basic exponential family distribution such as exponential or gamma distribution.

*Example 8 (Kidney Disease Progression):* We look back to Example 5. The overall waiting time MGF is given in the equation (9). The first thing to do is we need to use partial fraction expansion and do some

arrangements to simplify $M_T(s)$ into a more familiar form of MGF. The calculation is done as follows:

$$\begin{aligned}M_T(s) &= M_{2\lambda_0}(s)M_{\lambda_1}(s)\\ &= \left(\frac{2\lambda_0}{2\lambda_0-s}\right)\left(\frac{\lambda_1}{\lambda_1-s}\right) \quad \text{for } s<\min(2\lambda_0,\lambda_1)\\ &= \left(\frac{2\lambda_0\lambda_1}{\lambda_1-2\lambda_0}\right)\left(\frac{1}{2\lambda_0-s}\right)+\left(\frac{2\lambda_0\lambda_1}{2\lambda_0-\lambda_1}\right)\left(\frac{1}{\lambda_1-s}\right)\\ &= \left(\frac{2\lambda_0\lambda_1}{2\lambda_0-\lambda_1}\right)\left(\frac{\lambda_1}{\lambda_1-s}\right)-\left(\frac{\lambda_1}{2\lambda_0-\lambda_1}\right)\left(\frac{2\lambda_0}{2\lambda_0-s}\right). \quad (21)\end{aligned}$$

To convert these MGFs into density we need to consider the cases $2\lambda_0 > \lambda_1$, $2\lambda_0 < \lambda_1$ and $2\lambda_0 = \lambda_1$ separately.

Case 1, $2\lambda_0 > \lambda_1$:

$$\begin{aligned}f_T(t) &= \left(\frac{2\lambda_0}{2\lambda_0-\lambda_1}\right)\lambda_1 e^{-\lambda_1 t}-\left(\frac{\lambda_1}{2\lambda_0-\lambda_1}\right)(2\lambda_0)e^{-2\lambda_0 t}\\ &= \left(\frac{2\lambda_0\lambda_1}{2\lambda_0-\lambda_1}\right)\left(e^{-\lambda_1 t}-e^{-2\lambda_0 t}\right)\\ &\quad \text{for } t>0, \lambda_0>0, \lambda_1>0, 2\lambda_0>\lambda_1.\end{aligned}$$
(22)

Case 2, $2\lambda_0 < \lambda_1$:

$$\begin{aligned}f_T(t) &= \left(\frac{\lambda_1}{\lambda_1-2\lambda_0}\right)(2\lambda_0)e^{-2\lambda_0 t}-\left(\frac{(2\lambda_0)}{\lambda_1-2\lambda_0}\right)\lambda_1 e^{-\lambda_1 t}\\ &= \left(\frac{2\lambda_0\lambda_1}{\lambda_1-2\lambda_0}\right)\left(e^{-2\lambda_0 t}-e^{-\lambda_1 t}\right)\\ &\quad \text{for } t>0, \lambda_0>0, \lambda_1>0, 2\lambda_0<\lambda_1.\end{aligned}$$
(23)

Since (22) and (23) are identical, we can write

$$f_T(t) = \left(\frac{2\lambda_0\lambda_1}{2\lambda_0-\lambda_1}\right)\left(e^{-\lambda_1 t}-e^{-2\lambda_0 t}\right)$$

for $t>0, \lambda_0>0, \lambda_1>0, 2\lambda_0 \neq \lambda_1$.
(24)

For case 3, $2\lambda_0 = \lambda_1$ the MGF can be reduced to

$$M_T(s) = \left(\frac{\lambda_1}{\lambda_1-s}\right)\left(\frac{\lambda_1}{\lambda_1-s}\right) = \left(\frac{\lambda_1}{\lambda_1-s}\right)^2, \quad s<\lambda_1$$
(25)

so that

$$f_T(t) = \lambda_1^2 t e^{-\lambda_1 t}, \quad t>0, \lambda_1>0, \quad (26)$$

which is the density of gamma distribution with mean $2/\lambda_1$. Partial fraction expansion becomes a longer and more difficult process as the number of transitions grows. For a more complicated model, particularly those with some nonexponential waiting times, exact or analytic inversion of the MGF can be very tedious or even impossible.

## III. MARKOV JUMP PROCESSES

Continuous time Markov chains or Markov jump processes (MJP) are continuous time stochastic process $X = \{X_t, t \in \mathcal{T}\}$ with a discrete state space $S = \{k_0, k_1, \ldots, k_{n-1}, k_n\}$ that satisfy the Markov property, i.e.

$$\begin{aligned}&P(X_{t_n} = k_n | X_{t_0} = k_0, X_{t_1} = k_1, \ldots, X_{t_{n-1}} = k_{n-1})\\ &= P(X_{t_n} = k_n | X_{t_{n-1}} = k_{n-1}),\end{aligned}$$
(27)

For all $n \geq 1$, $k_0, k_1, \ldots, k_n \in S$ and any finite sequence $t_n > t_{n-1} > \cdots > t_1 > t_0 \geq 0$ of times in $\mathcal{T}$ such that $P(X_{t_0} = k_0, X_{t_1} = k_1, \ldots, X_{t_{n-1}} = k_{n-1}) > 0$. The *transition probability* form state $i$ to state $k$ in an MJP is given by $p_{ik}(s,t) = P(X_t = k | X_s = i)$. Assume that the size of the sample space, $d = |S|$, is Finite. We can create a $d*d$ transition matrices $\mathbf{P}(s,t)$ in which the $(i,k)$-th entry is the transition probability $p_{ik}(s,t) \geq 0$ with $\sum_{k \in S} p_{ik}(s,t) = 1$ for all $i \in S$ (the row sums equal 1). The transition matrices $\mathbf{P}(s,t)$ of an MJP also satisfy Chapman-Kolmogorov equations, i.e. $p_{ik}(s,t) = \sum_{j \in S} p_{ij}(s,u) p_{jk}(u,t)$ for all $i,k \in S$ and any $u$ with $s < u < t$, or

$$\mathbf{P}(s,t) = \mathbf{P}(s,u)\mathbf{P}(u,t). \quad (28)$$

A set of $d*d$ matrices $\{\mathbf{P}(s,t), s \geq 0, t \geq s\}$ satisfying the Chapman-Kolmogorov equations is called a transition matrix function. Transition matrix functions can be generated by a set of Q-matrices, where a Q-matrix or a *generator matrix* is a $d*d$ matrix $\mathbf{Q}(t) = (\mu_{ij}(t))_{i,j=1}^d$ with $\mu_{ij}(t) \geq 0$ for $i \neq j$ and $\sum_{j=1}^d \mu_{ij}(t) = 0$ for all $i = 1, \ldots, d$.

We can also say that the set of Q-matrices generates the Markov jump process X. The transition rates $\mu_{ij}(t)$ represent the instantaneous rate of change of $p_{ik}(s,t)$ at $t = s$, they are the most fundamental quantities in an MJP. We define $\mu_i(t) := -\mu_{ii}(t) = \sum_{j=1, j \neq i}^d \mu_{ij}(t)$ for the total transition rate out of state $i$, note that $\mu_i(t) \geq 0$.

The connection between transition probabilities and transition rates are explained using the so-called *Kolmogorov's differential equations*.

*Theorem 1 (Kolmogorov's differential equations):*
Let $\{\mathbf{P}(s,t), s \geq 0, t \geq s\}$ be a transition matrix function generated by the set of Q-matrices $\{\mathbf{Q}(t) = (\mu_{ij}(t))_{i,j=1}^d, t \geq 0\}$ where the entries $\mu_{ij}(t)$ are continuous and bounded in t. Then for each $t \geq 0$, $\mathbf{P}(s,t)$ satisfies the following systems of ODE:

1. Kolmogorov's forward differential equations

$$\frac{\partial}{\partial t}\mathbf{P}(s,t) = \mathbf{P}(s,t)\mathbf{Q}(t), \quad t>s, \quad (29)$$

i.e. for all $i,k \in S = \{1, \ldots, d\}$,

$$\frac{d}{dt} p_{ik}(s,t) = \sum_{j=1}^d p_{ij}(s,t)\mu_{jk}(t). \quad (30)$$

2. Kolmogorov's backward differential equations:

$$\frac{\partial}{\partial t}\mathbf{P}(s,t) = -\mathbf{Q}(s)\mathbf{P}(s,t), \quad t>s, \quad (31)$$

i.e. for all $i,k \in S = \{1, \ldots, d\}$,

$$\frac{d}{ds} p_{ik}(s,t) = -\sum_{j \in S} \mu_{ij}(s) p_{jk}(s,t). \quad (32)$$

With boundary condition $\mathbf{P}(s,s) = \mathbf{I}$.
A Markov jump process $X = \{X_t, t \geq 0\}$ generated by a set of Q-matrices $\{\mathbf{Q}(t) = (\mu_{ij}(t))_{i,j=1}^d, t \geq 0\}$ can be represented by a step function $t \mapsto X_t$, which actually forms a right-continuous sample path of the process. If $X$ is at state $i$ at time $s$, or $X_s = i$, the waiting time in state $i$ until the next jump is defined as a survival time with hazard function $t \mapsto \mu_i(s+t)$. The probability of $X$ jumps to state $k \neq i$ at time $t$ is given by $\mu_{ik}(t)/\mu_i(t)$.

### A. Time homogeneous MJP

A Markov chain is said to be *time homogeneous* if the transition probabilities $p_{ik}(s,t)$ depend only on the difference $(t-s)$ but not on the individual values of time $t$ and $s$, i.e. if $p_{ik}(s,t) = p_{ik}(0, t-s)$ for all $i, k \in S$. In a time homogeneous MJP, we can use the notation $p_{ik}(t) := p_{ik}(s, s+t) = p_{ik}(0,t)$ for the transition probability and $\mathbf{P}(t) := \mathbf{P}(s, s+t) = \mathbf{P}(0,t)$ for the transition matrix. In this case, the Equation (28) becomes
$$\mathbf{P}(s,t) = \mathbf{P}(s)\mathbf{P}(t).$$
A time homogeneous MJP is generated by the Q-matrix $\mathbf{Q} = (\mu_{ij})_{i,j=1}^d$. The transition rate out of state $I$ is denoted by $\mu_i := -\mu_{ii} = \sum_{j=1, j \neq i}^d \mu_j$. From the previous section, we know that if $X_s = i$, the waiting time in state $i$ has a hazard rate $\mu_i$. So in the time homogeneous case, that waiting time is exponentially distributed with parameter $\mu_i$ since it has a constant hazard rate. Moreover, the probability of the MJP jumps to state $k \neq i$, is $\mu_{ik}/\mu_i$. independent of the jump time.

## IV. FLOWGRAPH ANALYSIS FOR MARKOV PROCESSES

### A. Progression of cancer model

The flowgraph model of the progression of cancer has been described and analyzed in Example 3 and Example 6. Since the model has three states and the waiting times in each state follow exponential distribution, we can model it as a time homogeneous Markov jump process. Figure 22 below presents the multistate MJP model of the progression of cancer, with the branches labeled with the transition rates.

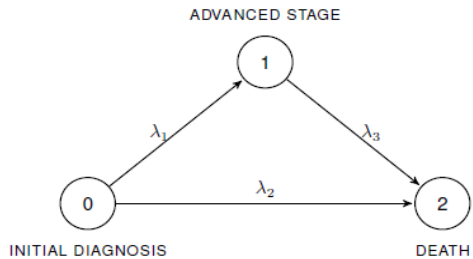

Figure 22. Multistate model for cancer progression

The state space of this model is $S = \{0, 1, 2\}$ with generator matrix
$$\mathbf{Q} = \begin{pmatrix} -(\lambda_1 + \lambda_2) & \lambda_1 & \lambda_2 \\ 0 & -\lambda_3 & \lambda_3 \\ 0 & 0 & 0 \end{pmatrix}$$
For simplicity, we furthermore assume that $\lambda_1 + \lambda_2 \neq \lambda_3$. The transition probabilities can be found by solving the Kolmogorov's forward differential equations for homogeneous case, $\frac{d}{dt}\mathbf{P}(t) = \mathbf{P}(t)\mathbf{Q}$ with boundary condition $\mathbf{P}(0) = \mathbf{I}$. First, we consider the first three forward equations corresponding to the backward state 0:

$$\frac{d}{dt}p_{00}(t) = -(\lambda_1 + \lambda_2)p_{00}(t)$$

$$\frac{d}{dt}p_{01}(t) = \lambda_1 p_{00}(t) - \lambda_3 p_{01}(t) \qquad (33)$$

$$\frac{d}{dt}p_{02}(t) = \lambda_2 p_{00}(t) + \lambda_3 p_{01}(t).$$

Another condition that must be satisfied is $p_{00}(t) + p_{01}(t) + p_{02}(t) = 1$ with the boundary conditions $p_{00}(0) = 1$, $p_{01}(0) = 0$, and $p_{02}(0) = 0$. If we calculate the differential equation in the first equation, it will give us $p_{00}(t) = Ae^{-(\lambda_1+\lambda_2)t}$ for some constant A. Since $p_{00}(t) = 1$, we get $A = 1$ and therefore $p_{00}(t) = e^{-(\lambda_1+\lambda_2)t}$. Substituting this result to the second equation, we must have $\frac{d}{dt}p_{01}(t) = \lambda_1 e^{-(\lambda_1+\lambda_2)t} - \lambda_3 p_{01}(t)$. Solving the homogeneous part of the differential equation will give us $p_{01}^{\text{hom}}(t) = Be^{-\lambda_3 t}$ for some constant B. Now let $p_{01}^{\text{part}}(t) = Ce^{-(\lambda_1+\lambda_2)t}$ be the particular solution to the nonhomogeneous ODE. Substituting this to the second equation will give is $C = \frac{\lambda_1}{\lambda_3 - \lambda_1 - \lambda_2}$. The general solution to the ODE is then calculated as

$$p_{01}(t) = p_{01}^{\text{hom}}(t) + p_{01}^{\text{part}}(t)$$
$$= Be^{-\lambda_3 t} + \frac{\lambda_1}{\lambda_3 - \lambda_1 - \lambda_2}e^{-(\lambda_1+\lambda_2)t}.$$

The boundary condition $p_{01}(0) = 0$ will gives us $B = -\frac{\lambda_1}{\lambda_3 - \lambda_1 - \lambda_2}$. Finally, the condition $p_{00}(t) + p_{01}(t) + p_{02}(t) = 1$ leads us to $p_{02}(t)$ since we already knew $p_{00}(t)$ and $p_{01}(t)$. Hence, the first three transition probabilities are

$$p_{00}(t) = e^{-(\lambda_1+\lambda_2)t}$$
$$p_{01}(t) = \frac{\lambda_1}{\lambda_1 + \lambda_2 - \lambda_3}(e^{-\lambda_3 t} - e^{-(\lambda_1+\lambda_2)t})$$
$$p_{02}(t) = 1 - \frac{\lambda_1}{\lambda_1+\lambda_2-\lambda_3}e^{-\lambda_3 t} + \frac{\lambda_3-\lambda_2}{\lambda_1+\lambda_2-\lambda_3}e^{-(\lambda_1+\lambda_2)t}.$$
(34)

The next three differential equations of the backward state 1 are

$$\frac{d}{dt}p_{10}(t) = -(\lambda_1 + \lambda_2)p_{10}(t)$$

$$\frac{d}{dt}p_{11}(t) = \lambda_1 p_{10}(t) - \lambda_3 p_{11}(t) \quad (35)$$

$$\frac{d}{dt}p_{12}(t) = \lambda_2 p_{10}(t) + \lambda_3 p_{11}(t).$$

with additional equation $p_{10}(t) + p_{11}(t) + p_{12}(t) = 1$ and boundary conditions $p_{10}(0) = 0$, $p_{11}(0) = 1$, and $p_{12}(0) = 0$. We can solve this with a similar procedure as before and get the transition probabilities

$$\begin{aligned}p_{10}(t) &= 0\\ p_{11}(t) &= e^{-\lambda_3 t}\\ p_{12}(t) &= 1 - e^{-\lambda_3 t}\end{aligned} \quad (36)$$

Furthermore, since state 2 is an end state, we then have $p_{20}(t) = p_{21}(t) = 0$ and $p_{22}(t) = 1$.

We want to see whether it is possible to find the total waiting time distribution of a multistate model without using flowgraph analysis or working on the MGF domain. Let $T$ be the total waiting time from state 0 to state 2. Our aim is to find the distribution of $T$. Assume that $T_0$ is the waiting time in state 0 and $T_1$ is the waiting time in state 1. From the previous section, we know that $T_0$ is exponentially distributed with parameter $\lambda_1 + \lambda_2$ and $T_1$ follows exponential distribution with parameter $\lambda_3$. Let $J(n), n = 1, 2, 3, ...$ be the $n$th jump time, the time of transition from one state to another state. The distribution of $T$ can be calculated as

$$\begin{aligned}P(T \leq t) &= P(T \leq t | X_{J(1)} = 1)P(X_{J(1)} = 1)\\ &\quad + P(T \leq t | X_{J(1)} = 2)P(X_{J(1)} = 2)\\ &= P(T_0 + T_1 \leq t)P(X_{J(1)} = 1)\\ &\quad + P(T_0 \leq t)P(X_{J(1)} = 2)\\ &= \left(1 + \frac{\lambda_3}{\lambda_1 + \lambda_2 - \lambda_3}e^{-(\lambda_1+\lambda_2)t}\right.\\ &\quad \left. - \frac{\lambda_1+\lambda_2}{\lambda_1+\lambda_2-\lambda_3}e^{-\lambda_3 t}\right)\left(\frac{\lambda_1}{\lambda_1+\lambda_2}\right)\\ &\quad + \left(1 - e^{-(\lambda_1+\lambda_2)t}\right)\left(\frac{\lambda_2}{\lambda_1+\lambda_2}\right)\\ &= 1 + \frac{\lambda_1\lambda_3 - \lambda_1\lambda_2 - \lambda_2^2 + \lambda_2\lambda_3}{(\lambda_1+\lambda_2)(\lambda_1+\lambda_2-\lambda_3)}e^{-(\lambda_1+\lambda_2)t}\\ &\quad - \frac{\lambda_1}{\lambda_1+\lambda_2-\lambda_3}e^{-\lambda_3 t}\end{aligned} \quad (37)$$

The cumulative distribution function of $T_0 + T_1$ can be calculated using convolution, as explained by Ogutunde, Odetunmibi, and Adejumo (2014). Furthermore, the density function of $T$ is then given by

$$\begin{aligned}f_T(t) &= \frac{d}{dt}P(T \leq t)\\ &= -\frac{(\lambda_1\lambda_3 - \lambda_1\lambda_2 - \lambda_2^2 + \lambda_2\lambda_3)(\lambda_1+\lambda_2)}{(\lambda_1+\lambda_2)(\lambda_1+\lambda_2-\lambda_3)}e^{-(\lambda_1+\lambda_2)t}\\ &\quad + \frac{\lambda_1\lambda_3}{\lambda_1+\lambda_2-\lambda_3}e^{-\lambda_3 t}\end{aligned} \quad (38)$$

We have calculated the MGF of the total waiting time from state 0 to state 2 of the cancer progression model in Example 6. It is given by

$$M_T(s) = \left(\frac{\lambda_1}{\lambda_1+\lambda_2}\right)\left(\frac{\lambda_1+\lambda_2}{\lambda_1+\lambda_2-s}\right)\left(\frac{\lambda_3}{\lambda_3-s}\right) + \left(\frac{\lambda_2}{\lambda_1+\lambda_2}\right)\left(\frac{\lambda_1+\lambda_2}{\lambda_1+\lambda_2-s}\right). \quad (39)$$

Exact inversion method can be used to compute the density of the total waiting time. Using partial fraction expansion and some rearrangement will give us

$$\begin{aligned}M_T(s) &= \left(\frac{\lambda_1}{\lambda_1+\lambda_2}\right)\left[-\left(\frac{(\lambda_1+\lambda_2)\lambda_3}{\lambda_1+\lambda_2-\lambda_3}\right)\left(\frac{1}{\lambda_1+\lambda_2-s}\right)\right.\\ &\quad \left.+ \left(\frac{(\lambda_1+\lambda_2)\lambda_3}{\lambda_1+\lambda_2-\lambda_3}\right)\left(\frac{1}{\lambda_3-s}\right)\right]\\ &\quad + \left(\frac{\lambda_2}{\lambda_1+\lambda_2}\right)\left(\frac{\lambda_1+\lambda_2}{\lambda_1+\lambda_2-s}\right)\\ &= -\left(\frac{\lambda_1\lambda_3}{\lambda_1+\lambda_2-\lambda_3}\right)\left(\frac{1}{\lambda_1+\lambda_2-s}\right)\\ &\quad + \left(\frac{\lambda_1\lambda_3}{\lambda_1+\lambda_2-\lambda_3}\right)\left(\frac{1}{\lambda_3-s}\right)\\ &\quad + \left(\frac{\lambda_2}{\lambda_1+\lambda_2-s}\right)\\ &= \left(\frac{-\lambda_1\lambda_3 + \lambda_1\lambda_2 + \lambda_2^2 - \lambda_2\lambda_3}{\lambda_1+\lambda_2-\lambda_3}\right)\left(\frac{1}{\lambda_1+\lambda_2-s}\right)\\ &\quad + \left(\frac{\lambda_1\lambda_3}{\lambda_1+\lambda_2-\lambda_3}\right)\left(\frac{1}{\lambda_3-s}\right)\\ &= -\frac{(\lambda_1\lambda_3 - \lambda_1\lambda_2 - \lambda_2^2 + \lambda_2\lambda_3)}{(\lambda_1+\lambda_2)(\lambda_1+\lambda_2-\lambda_3)}\left(\frac{\lambda_1+\lambda_2}{\lambda_1+\lambda_2-s}\right)\\ &\quad + \left(\frac{\lambda_1}{\lambda_1+\lambda_2-\lambda_3}\right)\left(\frac{\lambda_3}{\lambda_3-s}\right).\end{aligned}$$

Since we have got a known form of MGF, we can convert it into the corresponding density function:

$$\begin{aligned}f_T(s) &= -\frac{(\lambda_1\lambda_3 - \lambda_1\lambda_2 - \lambda_2^2 + \lambda_2\lambda_3)}{(\lambda_1+\lambda_2)(\lambda_1+\lambda_2-\lambda_3)}(\lambda_1+\lambda_2)e^{-(\lambda_1+\lambda_2)s}\\ &\quad + \left(\frac{\lambda_1}{\lambda_1+\lambda_2-\lambda_3}\right)\lambda_3 e^{-\lambda_3 s}.\end{aligned} \quad (40)$$

which is exactly the same as (38), the density function we obtained by the construction of MJP. Therefore, in this progression of cancer model, the distribution of the total waiting time can be computed without using flowgraph model.

## V. CONCLUSION

For a simple model like the progression of cancer model, flowgraph analysis does not really give any advantage in an analytical computation. However, in a more complex model, it would be too difficult if we want to find the total waiting time distribution by just analyzing the sample path. For example, if we have a model similar to Figure 22 but with the transition from state 1 to state 0 made possible, a loop (0→1→0) will be constructed and the computation of the total waiting time distribution will not be as straightforward as before because there will be infinitely many possible jumps combination from state 0 to state 2.

Therefore, it is much easier to work in MGF domain using flowgraph analysis to compute the distribution of the total waiting time.

An example where flowgraph could give an advantage in the analytical computation is the so-called "Birth and death process". Birth and death process is one example of Markov jump processes which have many applications in queuing theory, engineering, biology, and the financial sector. The process can be split into pure birth process which starts from state 0 of no birth and progress to the next states sequentially, until state $N$, and the pure death process which begins from state $N$, and progress continuously through the states $N-1, N-2, \ldots, 0$ until extinction. The flowgraph model of the combined birth and death process is a series model, with waiting time between the states exponentially distributed. Calculating the distribution of the total waiting time would be very difficult by just constructing a Markov jump process, since there are lots of feedback loops, and we have to use the so-called "Mason's rule" to calculate the MGF of the total waiting time.

Another advantage of the flowgraphs models is the analysis of the semi-Markov process. Semi-Markov process is the extension of Markov process in which the jump time to the next state depends on the current holding time in the present state. Although semi-Markov models have many applications in medical, biology, engineering, financial sector, and especially actuarial models, the data analysis can be very complicated and practical solutions are difficult to implement. Flowgraphs can model semi-Markov process and help for the analytical computation and data analysis. Also, we can then use a different type of distributions to make the model more realistic. Huzurbazar (2005) gives some examples of flowgraph analysis for semi-Markov processes. Warr and Collins (2014) have developed a more straightforward and practical method of solving for quantities of interest in semi-Markov processes by using flowgraph models as the basic element. Since there are actuarial models that work in a semi-Markov environment (Janssen and Manca, 2002), further research is recommended to solve real-world actuarial problems that can be modeled by semi-Markov processes using flowgraph analysis.

ACKNOWLEDGMENT

We thank Dr. Ronnie Loeffen for the helpful comments, and taking the time and effort to supervise the full version of this project.